\documentclass[final,hidelinks,onefignum,onetabnum]{siamart220329}
\usepackage{graphicx}
\usepackage{amsfonts}
\usepackage{matlab-prettifier}
\usepackage[nocompress]{cite}
\usepackage{upquote}
\usepackage{verbatim}
\usepackage[caption=false]{subfig}

\title{Computing Stokes flows in periodic channels via rational approximation\thanks{Submitted to the editors DATE.
\funding{YX would like to thank financial support from the UK EPSRC (EP/W522582/1).}} 
}
\author{Yidan Xue\thanks{School of Mathematics, Cardiff University, Cardiff CF24 4AG, UK (\email{XueY25@cardiff.ac.uk})}
}

\begin{document}

\maketitle

\begin{abstract}
Rational approximation has proven to be a powerful method for solving two-dimensional (2D) fluid problems. At small Reynolds numbers, 2D Stokes flows can be represented by two analytic functions, known as Goursat functions. Xue, Waters and Trefethen [\textit{SIAM J.~Sci.~Comput.}, 46 (2024), pp.~A1214--A1234] recently introduced the LARS algorithm (Lightning-AAA Rational Stokes) for computing 2D Stokes flows in general domains by approximating the Goursat functions using rational functions. In this paper, we introduce a new algorithm for computing 2D Stokes flows in periodic channels using trigonometric rational functions, with poles placed via the AAA-LS algorithm [Costa and Trefethen, \textit{European Congr.~Math.}, 2023] in a conformal map of the domain boundary. We apply the algorithm to Poiseuille and Couette problems between various periodic channel geometries, where solutions are computed to at least 6-digit accuracy in less than 1 second. The applicability of the algorithm is highlighted in the computation of the dynamics of fluid particles in unsteady Couette flows.
\end{abstract} 



\section{Introduction}

In the standard Cartesian coordinates $(x,y)$, the steady-state flow of an incompressible Newtonian fluid at low Reynolds number is described by the two-dimensional (2D) Stokes equations
\begin{align}
    \mu\nabla^2\mathbf{u}=\nabla{p}, \label{eq:stokes1} \\
    \nabla\cdot\mathbf{u}=0, \label{eq:stokes2}
\end{align}
where $\mathbf{u}=(u,v)^T$ is the 2D velocity field, $\nabla=(\partial/\partial{x},\partial/\partial{y})$, $p$ is the pressure, and $\mu$ is the viscosity. Since the flow is 2D and incompressible, a stream function $\psi$ can be defined by
\begin{equation}
    u=\frac{\partial\psi}{\partial{y}},\ v=-\frac{\partial\psi}{\partial{x}}.
\end{equation}
The stream function satisfies the biharmonic equation
\begin{equation}
    {\nabla}^4\psi=0.
\label{eq:biharmonic}
\end{equation}

We now consider 2D Stokes problems in $\Omega$ bounded by periodic channels, which are $l$-periodic in the $x$ direction. Hence the physical quantities should satisfy periodic conditions
\begin{equation}
u(x+l,y)=u(x,y),\ v(x+l,y)=v(x,y),\
p(x+l,y)=p(x,y)-\Delta{p}, \label{eq:periodic}
\end{equation}
for any $(x,y)$ in $\Omega$, where $\Delta{p}$ is the pressure drop over $l$ in $x$. The simplest example of pressure-driven periodic 2D Stokes flow is Poiseuille flow in a 2D straight channel.

\begin{figure}[h]
  \centering
  \includegraphics[width=.7\textwidth]{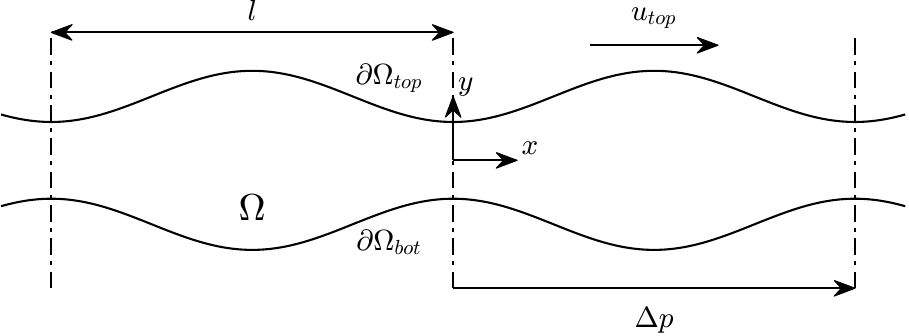}
  \caption{Schematic of 2D Stokes flow through a periodic channel.}
  \label{fig:schematic}
\end{figure}

\Cref{fig:schematic} shows a typical periodic 2D channel bounded by two no-slip periodic walls ($\partial\Omega_{top}$ and $\partial\Omega_{bot}$). In the conceptually most straightforward configuration, which we will call the ``Poiseuille problem'', these walls are stationary and the flow is steady, independent of time.  Such a flow is driven by a pressure gradient associated with a fixed pressure drop $\Delta{p}=p(x,y)-p(x+l,y)$, which will be independent of the choice of $(x,y)$. However, a central feature of this article is that we also consider configurations where the boundaries move steadily with time $t$, which we call the ``Couette problem''. Without loss of generality, we take the lower boundary to be stationary and allow the upper boundary to translate to the right with a fixed velocity $u_{top}$. In such a case, assuming the upper boundary is not simply a straight line, the variables $p$, $u$, $v$ and $\psi$ will all depend on $t$. However, the nature of Stokes flow is that each instant of time is uncoupled from the others, so that these quantities are all determined at an instant of time independently of the others. With this in mind, we will continue to use the notations $p(x,y)$, $u(x,y)$, $v(x,y)$ and $\psi(x,y)$ without noting the dependence on $t$ explicitly. In the Couette case, we assume $\Delta{p}=0$.

A central theme of the results we present is that the flows in the Couette case are unsteady, even though the boundary is translating at a steady speed. Our results focus therefore on pathlines, showing trajectories of particles of fluid, rather than streamlines, showing instantaneous flow directions. The pathlines we elucidate are complicated and sometimes chaotic.

Previous work on the Poiseuille problem primarily focuses on understanding mass transport and flow resistance in various boundary geometries. Assuming small boundary amplitude, the Poiseuille problem has been approximated by a series expansion of the stream function in symmetric sinusoidal boundaries \cite{Burns1967}, sinusoidal boundaries with a phase shift \cite{Wang1979}, a plane and an uneven wall \cite{Hasegawa1983} and more general boundaries \cite{Phan-Thien1980}. The Poiseuille problem has also been studied using numerical \cite{Sobey1980,Pozrikidis1987,Hemmat1995,Malevich2006,Bystricky2021} and experimental methods \cite{Stephanoff1980,Nishimura1984}. We note that most of these works only consider problems with smooth boundaries, except for Bystricky et al.~\cite{Bystricky2021}, who considers boundaries with sharp corners using an integral equation method.

Most existing literature on the Couette problem relates to engineering applications in tribology \cite{Plouraboue2001,Letalleur2002,vanOdyck2003}. In the context of fluids, Pozrikidis \cite{Pozrikidis1987} has considered Stokes flows and their streamline patterns between a moving plate and a sinusoidal wall using a boundary integral method. A similar scenario has been investigated experimentally using a gravity-driven film flow setup \cite{Wierschem2003}. Despite previous work on the steady Couette case (i.e.~where the boundaries are straight), there appears to be no literature on particle dynamics in unsteady Couette problems. One possible reason is that such computation requires great accuracy and speed in obtaining the solution of each quasi-steady problem, and evaluating the velocity field at the particle location. 

It is well known that the periodic movement of domain boundaries can lead to chaotic motion of fluid particles \cite{Aref1984}, particularly in Stokes flow problems \cite{Aref1986,Chaiken1986,Stone1991,Krasnopolskaya1999}. This phenomenon, known as topological chaos, has been used to improve the mixing efficiency of stirring devices \cite{Boyland2000,Finn2001,Finn2003}. We anticipate that topological chaos also exists in unsteady Couette problems, where the top boundary translates at a constant speed.

Since the introduction of the AAA algorithm \cite{Nakatsukasa2018}, the lightning algorithm \cite{Gopal2019a}, and the AAA-LS algorithm (AAA-least squares) \cite{Costa2023}, rational approximation has been shown to be a powerful method in computing 2D fluid flows \cite{Baddoo2020,Brubeck2022,Xue2024,McKee2024,Xue2024bifurcation}. In previous work \cite{Brubeck2022,Xue2024}, it has been shown that 2D Stokes problems in general domains can be solved using rational functions to 6-digit accuracy in less than 1 second. Using the computed solution, the evaluation of any physical quantity at any given point is extremely efficient, taking only tens of microseconds. These advantages suggest that rational approximation can be a suitable tool to compute periodic Stokes flows, particularly for computing the particle trajectories in unsteady Couette problems.

In this paper, we first introduce dimensionless variables in \cref{sec:non_dimension}, and the Goursat representation for 2D Stokes flows \cite{Goursat1898} in \cref{sec:goursat}. In \cref{sec:rational_basis}, we present new trigonometric rational function bases for computing periodic 2D Stokes flows, following \cite{Crowdy2019,Baddoo2021,Trefethen2023b,Xue2024}. We then introduce an algorithm to place the poles of the trigonometric rational functions by applying the AAA-LS algorithm \cite{Costa2023} in a conformal map of the domain boundary. After a Vandermonde with Arnoldi (VA) orthogonalization \cite{Brubeck2021}, we compute the coefficients in trigonometric rational function bases by solving a least-squares problem. The numerical implementations and example \hbox{MATLAB} codes are given in \cref{sec:numerics}. We apply the algorithm to a variety of Poiseuille and Couette problems in \cref{sec:results}. Using Poincar\'{e} maps \cite{Poincare1899,Henon1964,Strogatz2015}, we investigate particle dynamics and chaotic mixing in Couette problems between general $2\pi$-periodic boundaries in \cref{sec:unsteady_couette}.

\section{Non-dimensionalisation}
\label{sec:non_dimension}

For Poiseuille problems, we scale distances by $l/2\pi$, velocities by $\Delta{p}l/4\pi\mu$, and pressure by $\Delta{p}/2\pi$ to derive the dimensionless Stokes equations
\begin{align}
    \nabla^2\mathbf{u}=\nabla{p}, \label{eq:nd1} \\
    \nabla\cdot\mathbf{u}=0. \label{eq:nd2}
\end{align}
The problem now has boundary conditions
\begin{align}
u&=0,\ v=0,\quad\mathrm{on}\ \partial\Omega_{top},\\
u&=0,\ v=0,\quad\mathrm{on}\ \partial\Omega_{bot},
\end{align}
and periodic conditions
\begin{equation}
u(x+2\pi,y)=u(x,y),\ v(x+2\pi,y)=v(x,y),\ p(x+2\pi,y)=p(x,y)-2\pi.
\label{eq:periodic_poiseuille}
\end{equation}

For Couette problems, we scale distances by $l/2\pi$, velocities by $u_{top}$, and pressure by $2\pi\mu{u}_{top}/l$. Despite the same dimensionless Stokes equations, we now have boundary conditions
\begin{align}
u&=1,\ v=0,\quad\mathrm{on}\ \partial\Omega_{top},\\
u&=0,\ v=0,\quad\mathrm{on}\ \partial\Omega_{bot},
\end{align}
and periodic conditions
\begin{equation}
u(x+2\pi,y)=u(x,y),\ v(x+2\pi,y)=v(x,y),\ p(x+2\pi,y)=p(x,y).
\label{eq:periodic_couette}
\end{equation}

\section{A complex variable method using the Goursat functions}
\label{sec:goursat}

In the complex plane $z=x+iy$, where $i=\sqrt{-1}$, we have
\begin{equation}
    \frac{\partial}{\partial{z}}=\frac{1}{2}\left(\frac{\partial}{\partial{x}}-i\frac{\partial}{\partial{y}}\right),\ \frac{\partial}{\partial{\bar{z}}}=\frac{1}{2}\left(\frac{\partial}{\partial{x}}+i\frac{\partial}{\partial{y}}\right),
\end{equation}
where $\bar{z}=x-iy$ is the complex conjugate of $z$. The biharmonic equation \cref{eq:biharmonic} can now be written in complex form as
\begin{equation}
\frac{\partial^4\psi}{\partial^2z\partial^2\bar{z}}=0,
\end{equation}
which has a solution
\begin{equation}
\psi(z,\bar{z})=\mathrm{Im}[\bar{z}f(z)+g(z)],
\end{equation}
where $f(z)$ and $g(z)$ are two analytic functions, known as Goursat functions \cite{Goursat1898}.

The flow velocity, pressure and vorticity in terms of the Goursat functions are
\begin{align}
u-iv = -\overline{f(z)}+\bar{z}f'(z)+g'(z), \label{eq:bc1} \\
p-i\omega=4f'(z), \label{eq:bc2}
\end{align}
where $\overline{f(z)}$ is the complex conjugate of $f(z)$, and $\omega = \partial{v}/\partial{x}-\partial{u}/\partial{y}$ is the dimensionless vorticity magnitude \cite{Brubeck2022}.

\section{Rational function bases for the Goursat functions}
\label{sec:rational_basis}

In the LARS algorithm (Lightning-AAA Rational Stokes) \cite{Xue2024}, we approximated the Goursat functions $f(z)$ and $g(z)$ using two rational functions, $\hat{f}(z)$ and $\hat{g}(z)$, which consist of poles that capture the singularities of the problem \cite{Nakatsukasa2018,Gopal2019a,Brubeck2022,Costa2023}, a polynomial \cite{Brubeck2021} and a finite Laurent series for each hole in the domain (with a corresponding logarithmic term) \cite{Axler1986,Trefethen2018}:
\begin{align}
\hat{f}(z)&=\sum_{j=1}^m\frac{a_j^f}{z-\beta_j}+\sum_{j=0}^n{b_j^f}z^j+\sum_{i=1}^p\sum_{j=1}^q{c_{ij}^f}(z-z_i)^{-j}+\sum_{i=1}^p{d_i^f}\log(z-z_i),
\label{eq:rational_fz}\\
\hat{g}(z)&=\sum_{j=1}^m\frac{a_j^g}{z-\beta_j}+\sum_{j=0}^n{b_j^g}z^j+\sum_{i=1}^p\sum_{j=1}^q{c_{ij}^g}(z-z_i)^{-j}+\sum_{i=1}^p{d_i^g}\log(z-z_i)\notag\\
&-\sum_{i=1}^p\overline{d_i^f}[(z-z_i)\log(z-z_i)-z],
\label{eq:rational_gz}
\end{align}
where $\beta$ are the poles, $z_i$ is a point in the $i$th hole, and $a$, $b$, $c$ and $d$ are unknown complex coefficients. After a Vandermonde with Arnoldi orthogonalization \cite{Brubeck2022} and imposing the boundary conditions using \cref{eq:bc1,eq:bc2}, we obtain a well-conditioned least-squares problem for the coefficients in the rational functions. The least-squares problem can be solved easily via the backslash command in \hbox{MATLAB}.

To compute periodic Stokes flow, one may use the above algorithm \cite{Xue2024} and impose periodic boundary conditions of velocities and stresses on a domain of one period in $x$ \cite{Barnett2018}. However, there are two limitations in this approach. First, it introduces additional degrees of freedom to satisfy the periodic boundary conditions, and thus makes the computation more complicated. Second, the Goursat functions approximated by this approach only works for one period, while the physical quantities in other periods can only be obtained after their coordinates being moved to the original period.

Hence it is desirable to find new rational function bases for the Goursat functions, so that all physical quantities, including $u$, $v$ and $p$, satisfy the $2\pi$-periodicity conditions. Crowdy and Luca \cite{Crowdy2019} considered Stokes flows due to periodic point singularities in a channel, using a trigonometric function basis of $\zeta=e^{iz}$ to approximate the Goursat functions. However, this methodology is only applicable to straight channel problems. For more general periodic domains, Trefethen \cite{Trefethen2023b} computed a Laplace problem in a periodic curved domain (bounded by a straight top wall and a sinusoidal bottom wall) using a trigonometric polynomial basis of $\zeta=e^{-iz}$ and trigonometric poles near the curved boundary placed by the AAA algorithm \cite{Nakatsukasa2018} on a larger interval (from $x=-\pi$ to $x=3\pi$). In addition to the AAA algorithm, there is a AAAtrig algorithm to approximate periodic Laplace problems and periodic 2D potential flows using rational functions with poles that are $2\pi$-periodic \cite{Baddoo2021}.

Following these previous works and \cite{Scholle2004a,Scholle2004b}, we define the form of the Goursat functions for 2D Stokes flows within general periodic boundaries:
\begin{align}
\hat{f}(z)&=-i\hat{a}z-3\hat{b}z^2+\tilde{f}(e^{iz}),\label{eq:fz_periodic}\\
\hat{g}(z)&=i\hat{a}z^2+\hat{b}z^3-z\tilde{f}(e^{iz})+\tilde{g}(e^{iz}),\label{eq:gz_periodic}
\end{align}
where $\hat{a}$ and $\hat{b}$ are real coefficients. Following the notations used in \cite{Crowdy2019}, $\tilde{f}(e^{iz})$ and $\tilde{g}(e^{iz})$ are two rational functions in the $\zeta=e^{iz}$ plane:
\begin{equation}
\tilde{r}(e^{iz})=\sum_{j=0}^k\frac{\hat{c}_j}{e^{iz}-e^{i\beta_j}}+\sum_{j=-m}^{n}\hat{d}_je^{ijz},
\label{eq:periodic_rational}
\end{equation}
which can be seen as rational functions in the $\zeta=e^{iz}$ plane consisting of poles $\zeta(\beta)$, a degree $m$ Laurent series about $\zeta=0$, and a degree $n$ polynomial. We will discuss methods to place poles in the next section.

From \cref{eq:fz_periodic,eq:gz_periodic}, we have
\begin{align}
u-iv&=-\overline{f(z)}+\overline{z}f'(z)+g'(z)\nonumber\\
&=-4\hat{a}\mathrm{Im}(z)-12\hat{b}\mathrm{Im}(z)^2-2\mathrm{Re}(\tilde{f}(e^{iz}))-2i\mathrm{Im}(z)(\tilde{f}(e^{iz}))'+(\tilde{g}(e^{iz}))'\label{eq:periodic_bc1}\\
p-i\omega&=4f'(z)\nonumber\\
&=-4i\hat{a}-24\hat{b}z+4(\tilde{f}(e^{-iz}))',\label{eq:periodic_bc2}\\
\psi&=\mathrm{Im}(\overline{z}f(z)+g(z))\nonumber\\
&=-2\hat{a}[\mathrm{Im}(z)]^2-4\hat{b}[\mathrm{Im}(z)]^3-2\mathrm{Im}(z)\mathrm{Re}(\tilde{f}(e^{iz}))+\mathrm{Im}(\tilde{g}(e^{iz})),\label{eq:periodic_bc3}
\end{align}
which are all $2\pi$-periodic in $x$.

We need to make a few comments before introducing the algorithm to approximate the Goursat functions. Firstly, all derivatives above are derivatives with respect to $z$. For Couette problems, $\hat{b}=0$. For Poiseuille problems, $\hat{b}=1/24$ from \cref{eq:periodic_poiseuille,eq:periodic_bc2}. Lastly, $\hat{f}(z)$ and $\hat{g}(z)$ cannot have other polynomial terms of $z$, or $\psi(z,\bar{z})=\mathrm{Im}[\bar{z}f(z)+g(z)]$ will have terms of $\mathrm{Re}(z)$ that are not periodic in $x$.

\section{Numerical scheme}
\label{sec:numerics}

Numerical schemes for computing 2D Stokes flows using rational functions have been introduced previously with example \hbox{MATLAB} codes in \cite{Brubeck2022,Xue2024}. In summary, we first sample points along the domain boundary and cluster poles near the singularities of the domain geometry using the lightning algorithm \cite{Gopal2019a} or the AAA algorithm \cite{Nakatsukasa2018}. After performing the AAA algorithm on the domain boundary to place poles, we remove the poles inside the boundary and keep the exterior poles to approximate the Goursat functions, following the AAA-LS algorithm for computing Laplace problems \cite{Costa2023}.

Next, we define a rational function basis for the Goursat functions consisting of a polynomial, partial fraction of poles, a Laurent series and a corresponding logarithmic term for each hole in the domain \cite{Xue2024}. We orthogonalise each basis for all sample points using the VA algorithm \cite{Brubeck2021} to obtain a matrix $A$, and impose boundary conditions on the sample points in a column vector $b$. Lastly, we compute the coefficient vector $x$ of the rational function by solving a least squares problem: $\min_x||Ax-b||$. In this section, we introduce a new numerical scheme for computing periodic 2D Stokes flows using rational function bases \cref{eq:fz_periodic,eq:gz_periodic}.

\subsection{AAA poles for periodic boundaries}
\label{sec:aaa_poles}

In our previous paper \cite{Xue2024}, we showed that the AAA poles near analytic boundaries are key to achieving rapid convergence when computing Stokes flows with smooth boundaries. The convergence rates of polynomial and rational approximations have also been discussed in more detail in \cite{Costa2023,Trefethen2023a,Trefethen2024}. The AAA poles can be obtained by approximating the Schwarz function \cite{Davis1958} on the analytic boundary by executing
\vspace{6pt}

\begin{small}
\begin{verbatim}
    F = conj(Zb);
    [r,pol] = aaa(F,Zb,'tol',1e-8,'mmax',200);
\end{verbatim}
\end{small}
\vspace{6pt}
where $Z_b$ is the vector of sample points along the curved boundary, near which we aim to place poles. We set the tolerance as $10^{-8}$ and the maximal degree as 200, in case the boundary geometry is complex. For most cases, the rational approximation degree is below 100. For truly complex boundaries, it is enough to rely on local AAA approximations of lower degree, as described in \cite{Costa2023}). Based on the AAA-LS algorithm \cite{Costa2023}, the next key step is that we only keep the poles outside the analytic regions. For example, when we compute Stokes flows or potential flows inside a domain $\Omega$, we remove the poles inside $\Omega$ and only keep the exterior poles:
\vspace{6pt}

\begin{small}
\begin{verbatim}
    inpoly = @(z,w) inpolygon(real(z),imag(z),real(w),imag(w));
    jj = inpoly(pol,Z);
    Pol = pol(~jj);
\end{verbatim}
\end{small}
\vspace{6pt}
where $Z$ is the vector of sample points along all of $\partial\Omega$. 

Here we extend existing methods to place poles for general periodic boundaries using the AAA algorithm \cite{Nakatsukasa2018,Costa2023}:
\begin{enumerate}
    \item Create a vector $Z_{top}$ of sample points along the top boundary $\partial\Omega_{top}$ and a vector $Z_{bot}$ along the bottom boundary $\partial\Omega_{bot}$, when $0\leq\mathrm{Re}(z)<2\pi$.
    \item Run the AAA algorithm to approximate the Schwarz function $F=\bar{z}$ on the conformal map of the boundary $\zeta_{top}=e^{iZ_{top}}$ and $\zeta_{bot}=e^{iZ_{bot}}$.
    \item Move the poles of the rational function back to the $z$-plane via $z=-i\log(\zeta)$.
    \item Move the poles into the interval $\mathrm{Re}(z)\in[0,2\pi)$. This step is necessary for \hbox{MATLAB}, because it uses $(-\pi,\pi]$ for the argument of its \texttt{log} function.
    \item Remove the poles inside the domain $\Omega$ and keep the exterior poles.
\end{enumerate}
We execute the following code in \hbox{MATLAB}:
\vspace{6pt}

\begin{small}
\begin{verbatim}
    inpolygonc = @(z,w)inpolygon(real(z),imag(z),real(w),imag(w)); 
    [r, pol1] = aaa(conj(zeta_top),zeta_top,'tol',1e-8,'degree',200);
    [r, pol2] = aaa(conj(zeta_bot),zeta_bot,'tol',1e-8,'degree',200);
    pol = [pol1;pol2];
    pol = -1i*log(pol);
    pol = [pol(real(pol)>=0); pol(real(pol)<0)+2*pi];
    ii = find(~inpolygonc(pol,Z) & real(pol)>=0 & real(pol)<2*pi);
    Pol = pol(ii);
\end{verbatim}
\end{small}
\vspace{6pt}

The AAAtrig algorithm \cite{Baddoo2021} performs rational approximation for functions that are $2\pi$-periodic in $x$ using cst trigonometric functions, as defined by Henrici in \cite{Henrici1979}. Through a transformation presented in Section 2.4 of \cite{Baddoo2021}, the trigonometric rational approximation is equivalent to a rational approximation in the $\zeta=e^{iz}$ plane, but with a scaled weight vector. Hence, the poles of the AAAtrig rational approximation can also be used to compute 2D Stokes flows in periodic boundaries. Numerical investigations with existing examples show that the poles obtained from the AAA algorithm (using the algorithm described above) and the AAAtrig algorithm have almost identical locations, resulting in computations of similar speed and accuracy. We will use the AAA algorithm for the rest of this paper.

We now consider two types of periodic geometries and present the AAA poles in both the $\zeta=e^{iz}$ and $z$ planes. In the first example, the domain boundaries are two analytic curves defined by: $y_1(x)=0.5+0.25\sin(4x+\pi/2)$ on $\partial\Omega_{top}$, and $y_2(x)=-y_1(x)$ on $\partial\Omega_{bot}$. In the second example, the domain boundaries are two piecewise functions with sharp corners:
\begin{equation}
  y_1(x) =
  \begin{cases}
    0.5 & \quad\mathrm{if}\ x\in[0,2\pi/3) \\
    0.5+1.5(x/\pi-2/3) & \quad\mathrm{if}\ x\in[2\pi/3,\pi) \\
    1 & \quad\mathrm{if}\ x\in[\pi,5\pi/3) \\
    1-1.5(x/\pi-5/3) & \quad\mathrm{if}\ x\in[5\pi/3,2\pi)
  \end{cases}\quad\mathrm{on}\ \partial\Omega_{top},
  \label{eq:polygonal_shape}
\end{equation}
and $y_2(x)=-y_1(x)$ on $\partial\Omega_{bot}$.

\begin{figure}[h!]
  \centering
  \includegraphics[width=\textwidth]{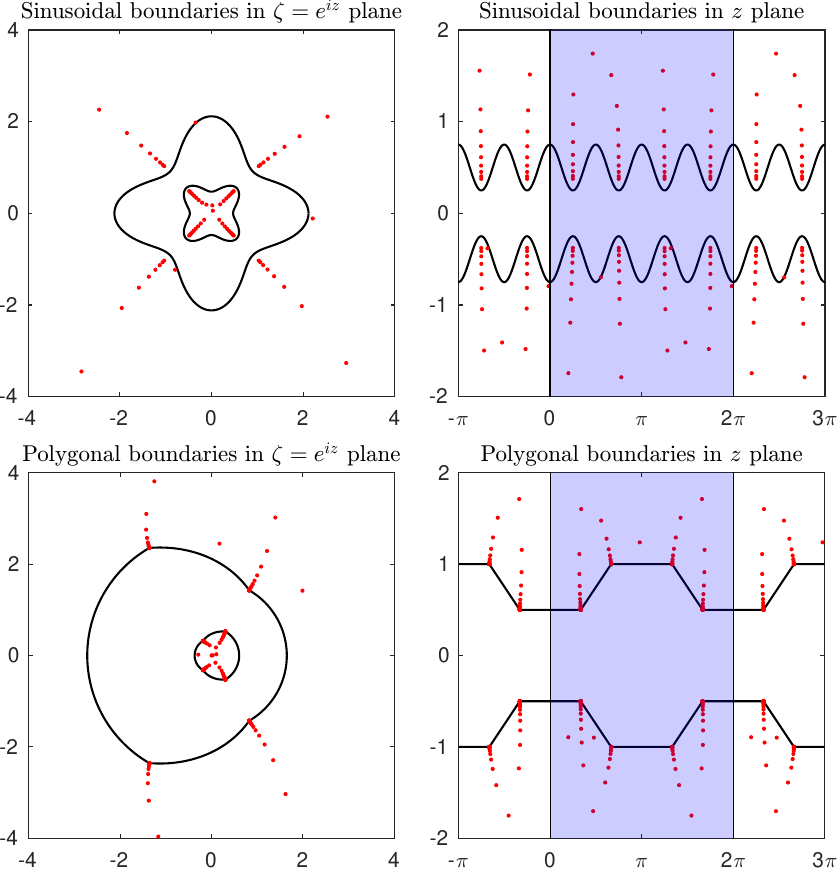}
  \caption{AAA poles outside the domain boundaries in the $\zeta=e^{iz}$ and $z$ planes. The poles are marked by red dots. In the top row, the poles are placed outside sinusoidal boundaries, which are periodic analytic curves. In the bottom row, the poles are placed outside polygonal boundaries, which are periodic piecewise functions with sharp corners. The AAA approximations are computed for the domain boundary in the blue region, where $\mathrm{Re}(z)\in[0,2\pi)$.}
  \label{fig:aaa_poles}
\end{figure}

\Cref{fig:aaa_poles} presents the poles placed by the AAA algorithm in $\bar{\Omega}$ after the conformal map $\zeta=e^{iz}$, shown as red dots in both $\zeta$ and $z$ planes. In the top row, the poles are placed outside the sinusoidal boundaries. Very similarly to the constricted channel case presented in \cite{Xue2024}, the poles cluster towards branch points of the analytic continuation across the boundary in the $z$-plane. In the bottom row, the poles cluster towards the sharp corners of the polygonal boundaries in the $z$-plane, which agrees with previous AAA and lightning approximations \cite{Gopal2019a,Brubeck2022,Costa2023}.
These poles will be used to capture the singularities of the geometries when approximating the Goursat functions.

\subsection{Vandermonde with Arnoldi process}

We construct a function basis consisting of the periodic AAA poles, the degree $m$ Laurent series, and the degree $n$ polynomial in \cref{eq:periodic_rational} using the VA orthogonalization \cite{Brubeck2021}. The orthogonalizations of the Laurent series and the polynomial have been presented in \cite{Xue2024} and \cite{Brubeck2022}, respectively. The orthogonalization of periodic poles only requires minor changes in \lstinline[style=Matlab-bw]{VAorthog} and \lstinline[style=Matlab-bw]{VAeval}, where \lstinline[style=Matlab-bw]{1./(Z-pol(k))} needs to be replaced with \lstinline[style=Matlab-bw]{1./(exp(1i*Z)-exp(1i*pol(k)))}. After the VA orthogonalization, we obtain a well-conditioned rational function basis $R_0$ for \cref{eq:periodic_rational} and its derivative $R_1$.

\subsection{Solving a linear least-squares problem}

We prescribe the movement $(u,v)$ of the top and bottom boundaries (using sample points $Z_{top}$ and $Z_{bot}$) to compute the coefficients in two rational functions $\hat{f}(z)$ and $\hat{g}(z)$ by solving a linear least-squares problem $\min_x||Ax-b||$. Using \cref{eq:periodic_bc1}, we construct the matrix $A$ and the vector $b$:
\begin{align}
A=&\left[
\begin{matrix}
-4\mathrm{Im}\{Z\} & -2\mathrm{Re}\{R_0\}+2\mathrm{Im}\{dZ\}\mathrm{Im}\{R_1\}\ & \mathrm{Re}\{R_1\}\\
0Z & 2\mathrm{Im}\{dZ\}\mathrm{Re}\{R_1\} & -\mathrm{Im}\{R_1\}
\end{matrix}
\right.\nonumber
\\
&\qquad\left.
\begin{matrix}
2\mathrm{Im}\{R_0\}+2\mathrm{Im}\{dZ\}\mathrm{Re}\{R_1\} & -\mathrm{Im}\{R_1\}\\
-2\mathrm{Im}\{dZ\}\mathrm{Im}\{R_1\} & -\mathrm{Re}\{R_1\}
\end{matrix}
\right],
b = \begin{bmatrix}
u(Z)+12\hat{b}\mathrm{Im}\{Z\}^2\\
v(Z)
\end{bmatrix}
\label{eq:linear}
\end{align}
where $0Z$ is a zero vector of the same size as $Z$ and $dZ=\mathrm{diag}(Z)$. The matrix $A$ has $2\times5$ blocks, where each column corresponds to $\hat{a}$, $\mathrm{Re}\{\hat{c}_j^f, \hat{d}_j^f\}$, $\mathrm{Re}\{\hat{c}_j^g, \hat{d}_j^g\}$, $\mathrm{Im}\{\hat{c}_j^f, \hat{d}_j^f\}$, $\mathrm{Im}\{\hat{c}_j^g, \hat{d}_j^g\}$, and each row corresponds to boundary conditions $u(Z)$ and $v(Z)$, respectively. We compute the least-squares problem using the backslash command in \hbox{MATLAB} to obtain the unknown coefficients in rational functions $\hat{f}(z)$ and $\hat{g}(z)$ for two Goursat functions.

\subsection{Computing the motion of fluid particles in unsteady problems}
\label{sec:4RK}

The unsteady Couette flow problem can be approximated as a series of independent quasi-steady problem, where the velocity field $uv(z,t)$ at time $t$ is computed using the Goursat functions, given by
\begin{equation}
uv(z,t) = -f(z,t)+z\overline{f'(z,t)}+\overline{g'(z,t)}.
\end{equation}
Assuming that the fluid particles are passively advected by the velocity field, their motions are governed by the variable-coefficient ordinary differential equation:
\begin{equation}
\frac{dz}{dt}=uv(z,t).
\end{equation}
For a particle $P$ initially at $P(t_0)=z_0$, we compute its location after a small interval $dt$ using a $4$th-order Runge-Kutta method via
\begin{align}
k_1 &= uv(z_0,t_0),\nonumber\\
k_2 &= uv(z_0+k_1dt/2,t_0+dt/2),\nonumber\\
k_3 &= uv(z_0+k_2dt/2,t_0+dt/2),\nonumber\\
k_4 &= uv(z_0+k_3dt,t_0+dt),\nonumber\\
P(t_0+dt) &= z_0+(k_1+2k_2+2k_3+k_4)dt/6.
\end{align}
To simulate the trajectories of fluid particles over multiples periods of $T=2\pi$, we compute the velocity field at $k$ equispaced snapshots for $dt=4\pi/k$ within a period, and use these solutions iteratively throughout the simulation. For the scenarios presented in this paper, $k=50$ is sufficient to achieve a reasonably accurate solution of the trajectory of a non-chaotic particle for hundreds of periods.

\section{Results}
\label{sec:results}

In this section, we apply the algorithm to a variety of 2D Stokes flow problems in periodic channels. The boundary conditions for these dimensionless problems were given in \cref{sec:non_dimension}.

\subsection{Poiseuille problems}

\begin{figure}[h!]
  \centering
  \includegraphics[width=\textwidth]{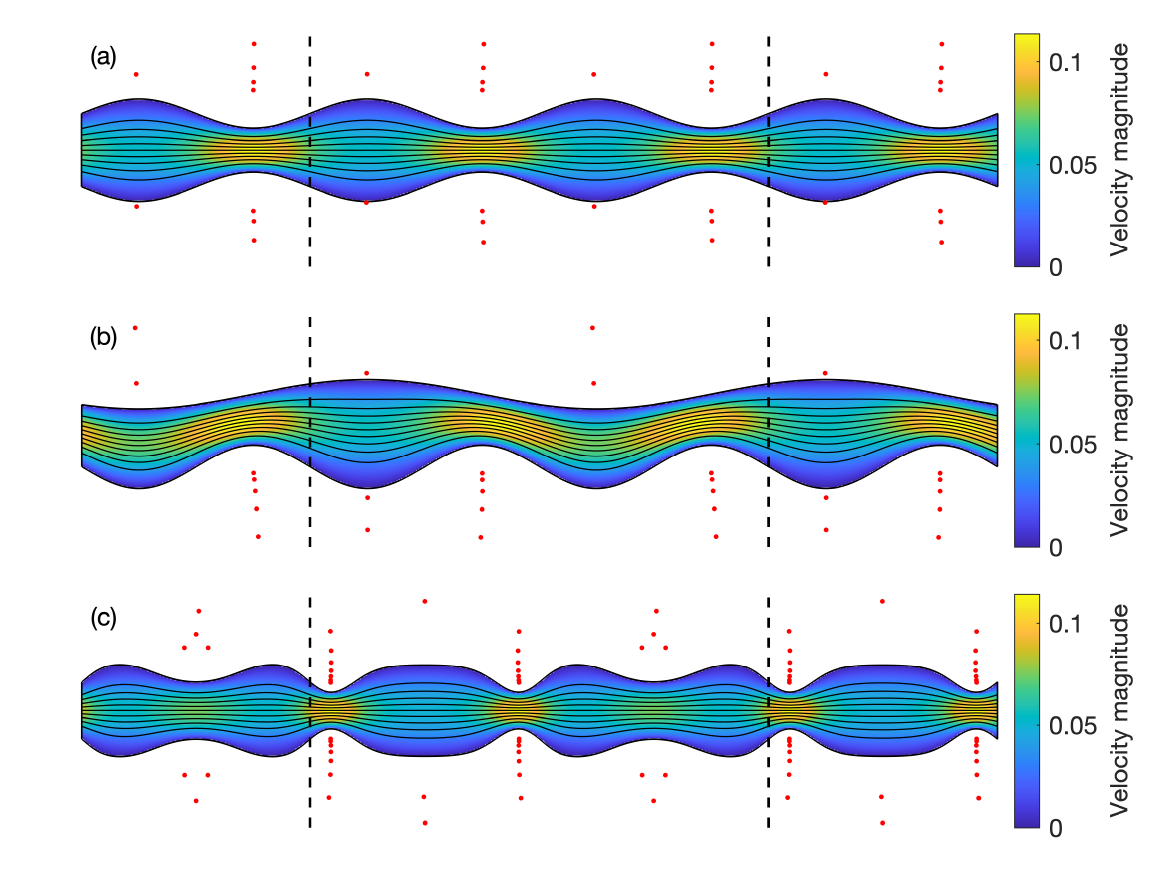}
  \caption{Stokes flows in three periodic channels with analytic boundaries. The streamlines, poles and velocity magnitude are represented by solid black lines, red dots and a colour scale. For each example, we present solutions from $x=-\pi$ to $x=3\pi$, where the period from $x=0$ to $x=2\pi$ is marked by two dashed black lines.}
  \label{fig:poiseuille_smooth}
\end{figure}

\Cref{fig:poiseuille_smooth} presents Stokes flows in three stationary periodic channels with analytic boundaries. The streamlines, poles and velocity magnitude are represented by solid black lines, red dots and a colour scale. A degree 15 finite Laurent series and a degree 15 polynomial are used for the rational function in the $\zeta$-plane (i.e.~$m=n=15$ in \cref{eq:periodic_rational}). Every case is computed to at least 6-digit accuracy in less than 0.2 second. One can obtain a 10-digit solution in a fraction of a second using the default tolerance of the AAA algorithm ($10^{-13}$ instead of $10^{-8}$).

In \cref{fig:poiseuille_smooth}(a), we compute Stokes flows in a symmetric channel bounded by two analytic curves: $y_1(x)=0.5+0.2\sin(2x)$ on $\partial\Omega_{top}$ and $y_2(x)=-y_1(x)$ on $\partial\Omega_{bot}$. In case (b), we consider flows in an asymmetric channel bounded by $y_1(x)=0.5+0.2\sin(x+\pi/4)$ on $\partial\Omega_{top}$ and $y_2(x)=-0.5-0.3\sin(2x)$ on $\partial\Omega_{bot}$. In case (c), we adapt a periodic function that has been treated by Fourier-based chebfuns \cite{Driscoll2014}: $f(x)=\tanh(\cos(1+2\sin(x))^2)-0.5$, where the top boundary is $y_1(x)=0.5+0.5f(x)$ and the bottom boundary is $y_2(x)=-y_1(x)$. We use this example to show that the AAA-LS algorithm \cite{Costa2023} is able to compute Stokes flows in complex periodic boundaries.

\begin{figure}[h!]
  \centering
  \includegraphics[width=\textwidth]{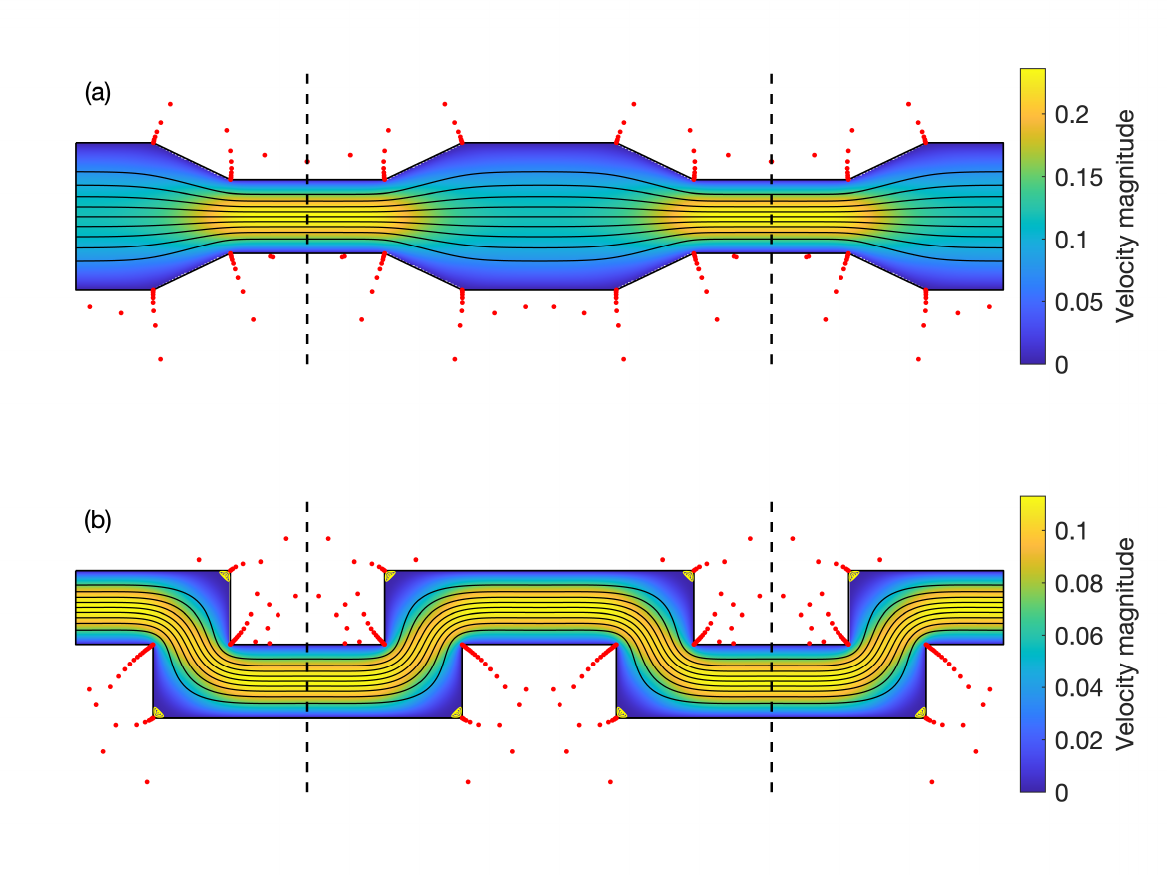}
  \caption{Stokes flows in two periodic channels with sharp corners. In (b), the eddies near sharp corners are shown using yellow contours of the stream function.}
  \label{fig:poiseuille_sharp}
\end{figure}

\Cref{fig:poiseuille_sharp} presents Stokes flows in two periodic channels with sharp corners. For \cref{fig:poiseuille_sharp}, the AAA-LS algorithm is run for nearby sample points for each corner singularity in the $\zeta$-plane to place the poles, following \cite{Costa2023}, to speed up the computation (i.e., the ``local AAA-LS'' method). To exploit the symmetry of the geometries, we find poles in $x\in[0,\pi]$ and flip those around $x=\pi$. \Cref{fig:poiseuille_sharp}(a) presents Stokes flows in the same polygonal boundaries as \cref{eq:polygonal_shape} and \cref{fig:aaa_poles}. In \cref{fig:poiseuille_sharp}(b), we consider Stokes flows in a periodic channel with right-angle corners. One can observe Moffatt eddies \cite{Moffatt1964} indicated by yellow contours near eight corners of $90^\circ$. The computation of two cases take 0.8 and 1.3 seconds, respectively.

\subsection{Steady Couette problems}

Unlike the Poiseuille problem, the Couette problem can be either steady or unsteady, depending on the boundary geometry. We will first consider a steady Couette problem, before investigating unsteady ones.

\begin{figure}[h]
  \centering
  \includegraphics[trim={3cm 0 3cm 0},clip,width=\textwidth]{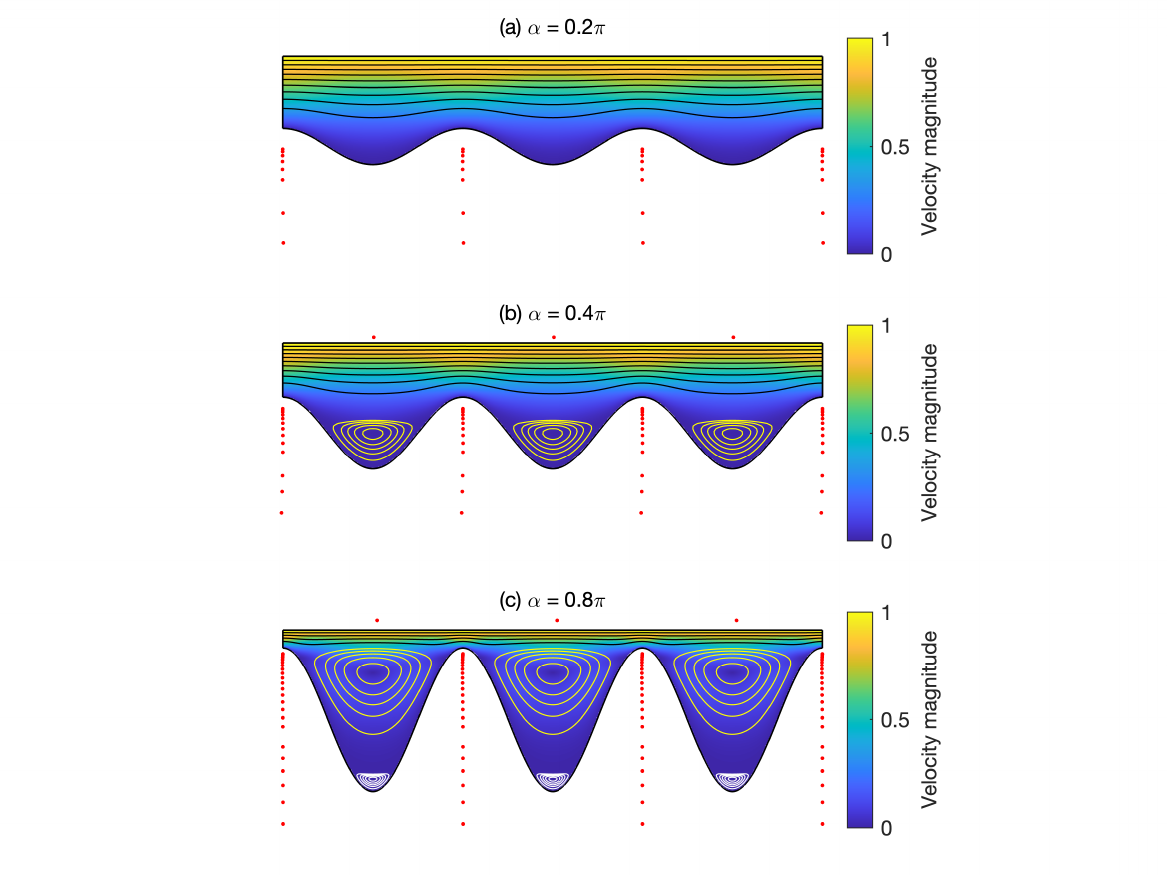}
  \caption{Stokes flows in periodic channels constricted by a moving flat wall and a steady sinusoidal wall. The first and second Moffatt eddies are indicated by yellow and white contours, respectively. The same problem has been computed in Figure $3$\emph{b} of \cite{Pozrikidis1987} using a boundary integral method.}
  \label{fig:steady_couette}
\end{figure}

\Cref{fig:steady_couette} presents Stokes flows through periodic channels constricted by a moving flat wall and a steady sinusoidal wall, following \cite{Pozrikidis1987}. The top boundary is $y_1(x) = \pi$ and the bottom boundary is $y_2(x) = \alpha\cos(x)$, where the amplitude $\alpha$ is $0.2\pi$, $0.4\pi$, and $0.8\pi$ for three subplots. The boundary conditions for one period from $x=0$ to $x=2\pi$ are used for computation, while we show the solutions in three periods from $x=-2\pi$ to $x=4\pi$.

For Couette problems, a degree 25 Laurent series and a degree 25 polynomial are used for the rational function in the $\zeta$-plane ($m=n=25$). The tolerance of the AAA algorithm is set as its default value of $10^{-13}$. Solutions for different values of $\alpha$ are obtained to more than 7-digit accuracy in less than 0.2 second.

When $\alpha=0.2\pi$, no eddy forms in the cavity regions. When $\alpha=0.4\pi$, one eddy forms in each cavity, the streamlines of which are indicated using yellow contours. When $\alpha=0.8\pi$, two eddies form in each cavity, where a primary eddy (yellow contours) occupies the upper half of the cavity and a secondary eddy (white contours) locates near the tip region. These agree with Pozrikidis' boundary integral simulations \cite{Pozrikidis1987}.

\subsection{Unsteady Couette problems}
\label{sec:unsteady_couette}

We now consider unsteady flows between two sinusoidal walls, which are described by $y_1(x)=1+\epsilon\sin(x-t)$ and $y_2(x)=-1+\epsilon\sin(x)$, respectively. The top boundary has a unit horizontal velocity, making the geometry of the problem time-periodic with a period $T=2\pi$. In this section, we examine how boundary geometry affects particle dynamics, focusing on their chaotic or non-chaotic behaviours.

We investigate the periodic problem using Poincar\'{e} maps, which capture the intersections of trajectories with a fixed cross-section \cite{Poincare1899,Henon1964,Strogatz2015}. Beginning with a set of particles $P_0$ in the spatial period between $x=0$ and $x=2\pi$ of $\Omega$, we simulate their movement until $t=nT=2n\pi$ for $n$ temporal periods using the $4$th-order Runge-Kutta method introduced in \cref{sec:4RK}. When $t=kT=2k\pi,\ k\in\mathbb{Z}^+$, we record particle locations as $P_k$. Finally, we plot all particle locations from $P_0,...,P_{n}$ superposed in the initial period via the mapping $x+iy\to\bmod(x,2\pi)+iy$. The Poincar\'{e} maps can be seen as a series of planes, $t=0,...,nT$, cutting through the $(2+1)$-dimensional space consisting of a 2D flow region and a time axis \cite{Boyland2000}. Following \cite{Henon1964}, the particles can exhibit two modes of behaviours on Poincar\'{e} maps: curves and clouds, which correspond to non-chaotic and chaotic regions, respectively.

A key characteristic of chaotic systems is their sensitive dependence on initial conditions. Starting with a point $P_0$ and a nearby point $P_0'$ separated by a small perturbation $\delta_0$, their trajectories will diverge exponentially, if the system is chaotic for initial condition $P_0$. Their gap $\delta_t$ can be described by
\begin{equation}
\|\delta_t\|\sim\|\delta_0\|e^{\lambda{t}},
\end{equation}
where $\lambda$ is known as the Lyapunov (or Liapunov) exponent \cite{Strogatz2015,Trefethen2017}. Despite being dependent on the initial condition $P_0$ and the perturbation $\delta_0$, a positive Lyapunov exponent is a signature of chaos.

To approximate the Lyapunov exponent, one may perform a least-squares fit of a linear equation to $\log|P_0-P_0'|,...,\log|P_n-P_n'|$ \cite{Trefethen2017}. Finn, Cox and Byrne \cite{Finn2003} used this method to compute the exponential stretching rate in a viscous mixer. However, for the scenarios considered in this paper, we observe that this method cannot always distinguish trajectories with slow (and sometimes oscillatory) exponential divergence from those with fast sub-exponential divergence. H\'{e}non and Heiles \cite{Henon1964} separated chaotic trajectories from non-chaotic ones by applying a threshold to the quantity $\rho=\sum_{i=1}^{25}|P_i-P_i'|^2$. This method amplifies the difference between the two types of behaviours using squares, and reduces the impact of oscillatory behaviours through the summation over successive periods. Following H\'{e}non and Heiles \cite{Henon1964}, we set an initial perturbation of $\|\delta_0\|=10^{-8}$ using $P_0'=P_0-10^{-8}i\cdot\mathrm{sgn}(\mathrm{Im}(P_0))$, or $P_0'=P_0+10^{-8}$ if $\mathrm{Im}(P_0)=0$, and determine a trajectory to be chaotic, if $\rho=\sum_{i=1}^{100}|P_i-P_i'|^2>10^{-7}$. This criterion, based on numerical experiments, can detect most $\delta_t$ with exponential growth in the cases considered in this paper.

\begin{figure}[h]
  \centering
  \includegraphics[width=\textwidth]{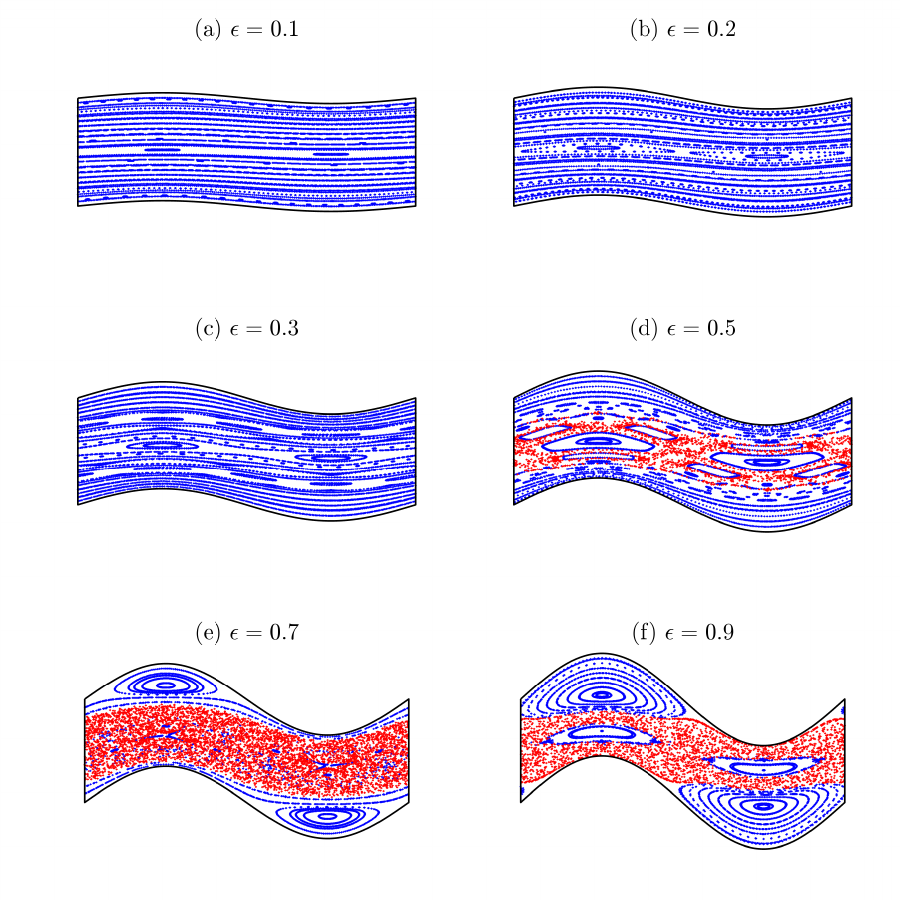}
  \caption{Poincar\'{e} maps of the unsteady Couette problem between two sinusoidal walls for different amplitude $\epsilon$. The locations of particle locations at integer multiples of the period $T$ until $t=300T$ are superposed in the initial period via the mapping $x+iy\to\bmod(x,2\pi)+iy$. The chaotic and non-chaotic particle trajectories are represented by red and blue dots, respectively.}
  \label{fig:poincare}
\end{figure}

\Cref{fig:poincare} shows the Poincar\'{e} maps of the unsteady Couette problem between sinusoidal walls for different amplitudes $\epsilon$. In these simulations, we begin with 40 particles, where 20 are evenly spaced along $x=\pi/2$ from the top boundary to the bottom boundary, and the other 20 are along $x=3\pi/2$. The simulation is terminated at $t=300T$, when the points have covered most of $\Omega$. For each particle, we determine whether its behaviour is chaotic using the criterion introduced before. On the Poincar\'{e} maps, we represent the chaotic and non-chaotic particles using red and blue dots, respectively. Agreements between the regions of particle clouds and those in red are shown in \Cref{fig:poincare}.

The behaviour of particle motion depends on its initial location and the amplitude $\epsilon$ of the sinusoidal boundary. For small $\epsilon$ (\cref{fig:poincare}a--c), the snapshots of each particle fall on a distinctive curve. On the Poincar\'{e} map, there are two stable points near $z=\pi/2$ and $z=3\pi/2$, where closed loops form around. No particle cloud is detected in \cref{fig:poincare}a--c, which indicates the particle motions are non-chaotic for small $\epsilon$. In \cref{fig:poincare}d, when $\epsilon=0.5$, particle clouds form around the closed loops, but bounded by curves near the domain boundary. For a larger $\epsilon=0.7$, eddies develop in the cavity region and particle clouds cover most of the central region. Despite their small size, a few blue islands can still be observed within the red particle cloud. Further increasing $\epsilon$ to 0.9 causes the expansion of the eddies in the cavity region, as well as the two closed loops, with most other regions covered by the clouds.

\begin{figure}[h]
  \centering
  \includegraphics[width=\textwidth]{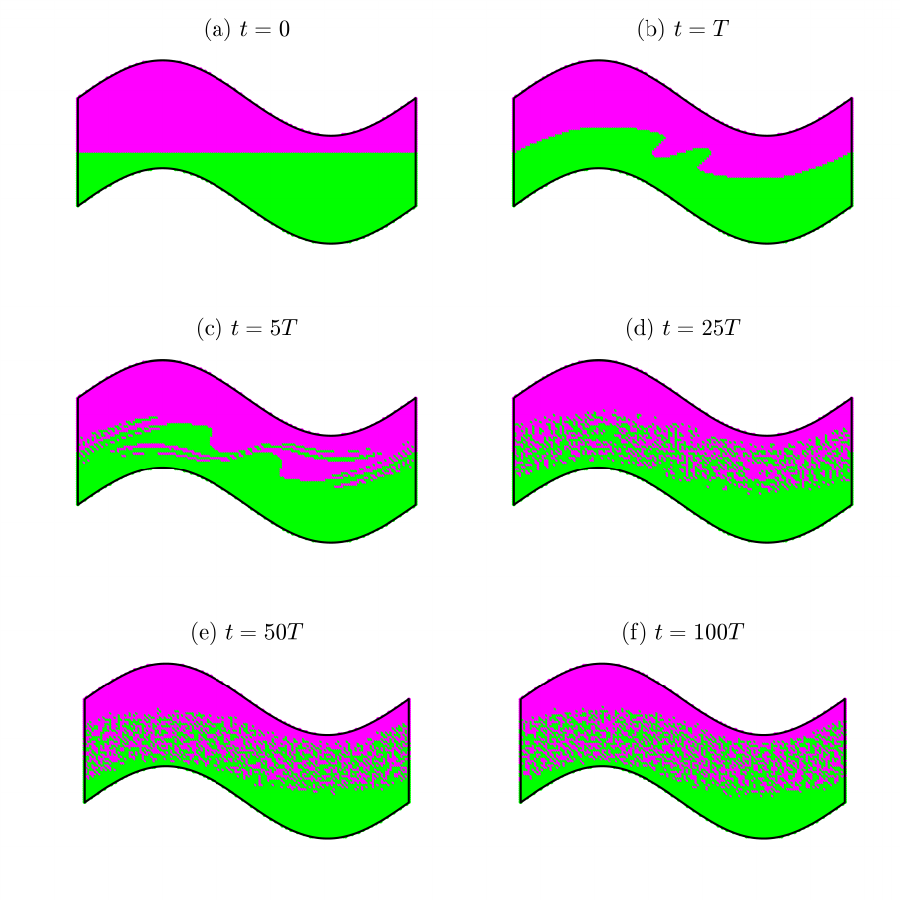}
  \caption{Mixing of particles in two colours between two sinusoidal walls for $\epsilon=0.7$ after different time intervals up to $t=100T$.}
  \label{fig:mixing}
\end{figure}

In \cref{fig:poincare}, the trajectories of chaotic particles fill an area densely, while those of the non-chaotic particles remain separated by gaps. This difference can result in significantly different mixing efficiencies in various regions, when diffusive effects are neglected \cite{Aref1986,Boyland2000,Finn2001,Finn2003}. \Cref{fig:mixing} presents the mixing of particles in two colours between two sinusoidal walls for $\epsilon=0.7$ after different time intervals. At $t=0$, all particles on the upper half with a positive imaginary part are magenta, while all particles on the lower half are green. To ensure consistent resolution for each plot, we start from an equispaced grid and run the simulation backwards (i.e.~reverse the boundary movements as described in \cref{sec:non_dimension}), thanks to the reversibility of Stokes flow. At each time $t$ for plotting, we check the imaginary part of each particle to determine its colour. We see from this computation that the mixing along vertical direction only happens in the chaotic regions as shown in \cref{fig:poincare}e. Similar results are observed in simulations for other $\epsilon$ values. Hence we may conclude that the mixing efficiency between two sinusoidal walls has a positive correlation with the fraction of chaotic regions in $\Omega$.

\begin{figure}[h]
  \centering
  \includegraphics[width=.7\textwidth]{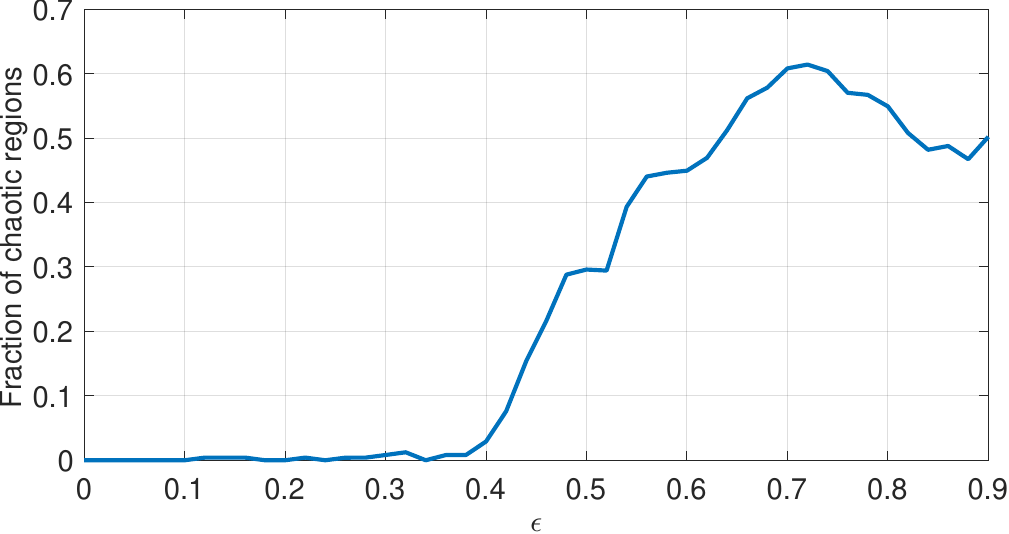}
  \caption{Fraction of chaotic regions in $\Omega$ for $\epsilon$ from $0$ to $0.9$.}
  \label{fig:chaotic_regions}
\end{figure}

To find $\epsilon$ that results in optimal mixing efficiency, we compute the fraction of chaotic regions for different $\epsilon$ from 0 to 0.9 with a step size of 0.02, as presented in \cref{fig:chaotic_regions}. We start with an equispaced particle grids in $\Omega$ and determine whether the trajectory of each particle is chaotic using the same method as used for \cref{fig:poincare}. The fraction of chaotic regions is estimated by calculating the ratio of particles exhibiting chaotic motion to the total number of particles. The maximum fraction of chaotic regions is found to be 0.614 for a ``critical amplitude'' of $\epsilon=0.72$, indicating the optimal mixing efficiency between sinusoidal walls.

\begin{figure}[h]
\centering 
\subfloat[Two polygonal boundaries]{\label{fig:poincare_case_a}\includegraphics[width=.6\linewidth]{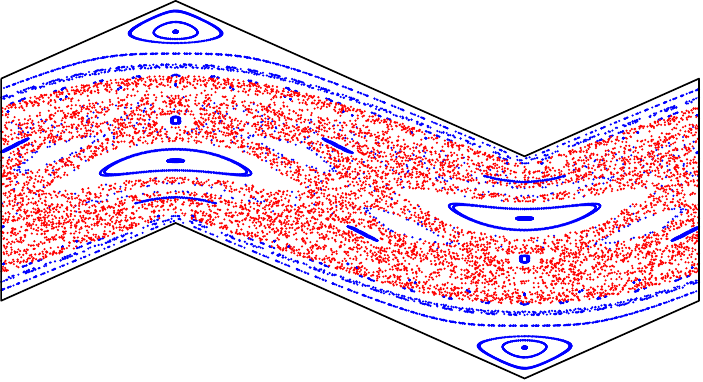}}\\ 
\subfloat[A polygonal boundary and a smooth boundary]{\label{fig:poincare_case_b}\includegraphics[width=.6\linewidth]{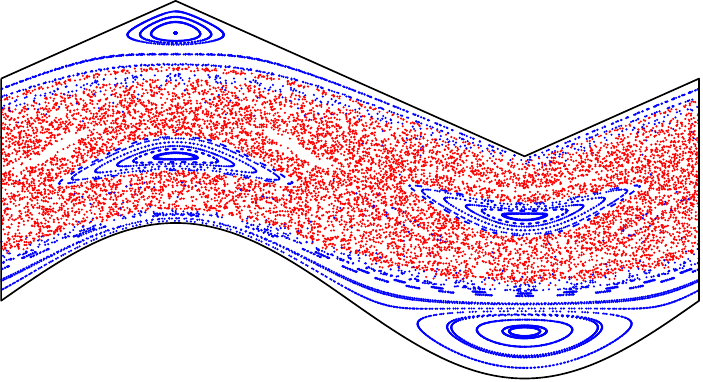}}\\
\subfloat[Two polygonal boundaries using the lightning poles]{\label{fig:poincare_case_c}\includegraphics[width=.6\linewidth]{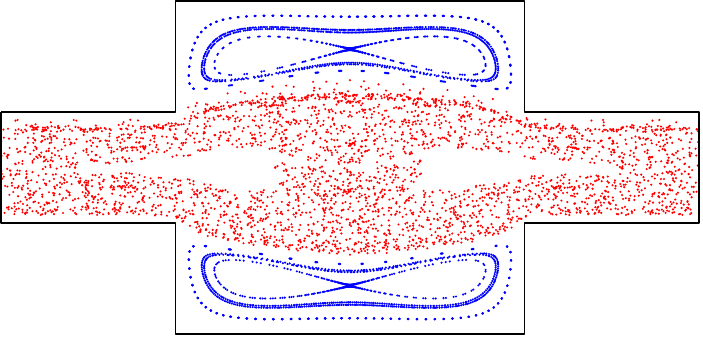}}\\
\caption{Poincar\'{e} maps of the unsteady Couette problem between different domain boundaries, generated by the same method as used for \cref{{fig:poincare}}. Cases (a) and (b) use the poles placed by the AAA algorithm, while case (c) uses the poles placed by the lightning algorithm.}
\label{fig:poincare_cases}
\end{figure}

Last, we apply the algorithm to compute Couette flows in other boundary geometries. \Cref{fig:poincare_case_a} presents the Poincar\'{e} maps for Couette flows between two polygonal boundaries computed by the same method as used for \cref{fig:poincare}. \Cref{fig:poincare_case_b} shows the Poincar\'{e} maps for Couette flows between a polygonal boundary and a smooth boundary. In \cref{fig:poincare_case_c}, we cluster 24 poles towards each sharp corner of the domain boundary based on the lightning algorithm, rather than placing the poles automatically using the AAA algorithm. The lightning poles also enable fast and accurate computations in periodic polygonal geometries.

\section{Discussion}

In this paper, we have presented an algorithm to compute Stokes flows in periodic channels using trigonometric rational functions. We map the domain boundary via $\zeta=e^{iz}$ and place the poles in the $\zeta$-plane by approximating the Schwarz function using the AAA algorithm \cite{Nakatsukasa2018}. Following the AAA-LS algorithm \cite{Costa2023}, the interior poles are removed and the exterior poles are used for the rational functions. We construct well-conditioned rational function bases for the Goursat functions using Vandermonde with Arnoldi orthogonalization \cite{Brubeck2021}, and approximate the coefficients in the rational functions by solving a least squares problem.

The algorithm presented inherits most advantages that have been shown in the lightning Stokes solver \cite{Brubeck2022} and the LARS algorithm \cite{Xue2024}, including a 6-digit accurate solution in 1 second and the evaluation time in tens of microseconds. In addition, the $2\pi$-periodicity condition is now satisfied by the rational function bases, making the computation of such problems much easier and faster than before. We showcase the applicability of the algorithm by computing a variety of periodic Stokes problems in \cref{sec:results}, including time-dependent problems that have not been considered in \cite{Xue2024}, and the trajectories of particle movements, which have never been considered between periodic channels with relative movements. 

As discussed in \cref{sec:aaa_poles}, we note that the algorithm presented is not the only way to compute periodic Laplace and Stokes problems using rational approximation. A similar algorithm can be built upon the AAAtrig algorithm \cite{Baddoo2021} and the AAA-LS algorithm \cite{Costa2023}, which places poles for the periodic boundaries using AAAtrig, and uses exterior poles to perform rational approximations following AAA-LS. Based on our numerical investigations of Stokes problems, no significant difference between the algorithm presented and the AAAtrig-least squares algorithm has been observed, regarding both speed and accuracy. However, we cannot rule out the possibility that one algorithm works better for certain problems than the other.

In this work, we have only considered periodic Stokes flows in simply connected domains. In \cite{Xue2024}, we extended the lightning Stokes solver \cite{Brubeck2022} to multiply connected domains by adding a Laurent series with a logarithmic term \cite{Trefethen2018} in each Goursat function, based on the the logarithmic conjugation theorem \cite{Axler1986}. However, applying a similar series method to periodic Stokes problems in multiply connected domains requires careful consideration of the rational function bases for the Goursat functions, since they need to satisfy the $2\pi$-periodicity conditions while ensuring the physical quantities are not multivalued. One way to compute periodic Stokes flows in multiply connected domains is to use the LARS algorithm \cite{Xue2024} with periodic boundary conditions \cite{Nicholls-Mindlin2023}. One can also compute such problems using the well-established integral equation methods \cite{Greengard2004,Barnett2018}.

In summary, we have presented an algorithm to compute periodic 2D Stokes flows using trigonometric rational functions and the AAA-LS algorithm. The computation takes less than a second to compute a solution with at least 6-digit accuracy. We have demonstrated its broad applicability by computing various Poiseuille and Couette flow problems, highlighted by the dynamics of fluid particles in unsteady scenarios.  

\section*{Acknowledgments}
I am indebted to Nick Trefethen for his encouragement and insightful suggestions that greatly improved this work. I am also grateful for helpful discussions with Jon Chapman, Stefan Llewellyn Smith, Kyle McKee, Sarah Waters, and Thomas Woolley.

\bibliographystyle{siamplain}
\bibliography{references}
\end{document}